\documentclass[a4paper,aps,twocolumn,superscriptaddress, showpacs,showkeys,floatfix]{revtex4}
\input{epsf}
\usepackage{graphicx}
\usepackage{epsfig}

\begin{document}

\title{
%Quantum FRW Cosmological Model With Matter and Cosmological Constant
Can Quantum Gravitational Effects Influence the Entire History of the Universe?}
\author{Ed\'esio M. Barboza Jr. \footnote{E-mail: edesio@on.br}}
\address{Departamento de Astronomia, Observat\'orio Nacional, 20921-400, Rio de Janeiro, Brazil}
\author{ Nivaldo A. Lemos \footnote{E-mail: nivaldo@if.uff.br}}
\address{Instituto de F\'{\i}sica, Universidade Federal Fluminense, 24210-340, Niter\'oi, Brazil}
\date{\today}
\begin{abstract}
In this work, a flat Friedmann-Robertson-Walker (FRW) universe with dust and a cosmological constant is quantized. By means of a canonical transformation, the classical Hamiltonian is reduced to that of either a harmonic oscillator or  anti-oscillator, depending on whether $\Lambda<0$ or $\Lambda>0$, respectively. In this way exact solutions to the Wheeler-DeWitt equation can easily be obtained. It turns out that a  positive  cosmological constant alone may account for an early inflationary regime and a later accelerated expansion phase, with  a period of decelerated expansion in between. This
suggests that  quantum gravitational effects can influence most of the history or even the entire history of the Universe.

\end{abstract}

\maketitle
\section{Introduction}

Recent observational data indicate that the Universe is going through a phase of accelerated expansion \cite{Riess, Perlmutter}. One of the main ideas  to explain this unexpected discovery consists in assuming that the the Universe is uniformly filled with a mysterious fluid, of unknown nature, called  dark energy. This exotic component, with equation of state $p/\rho=\textit{\texttt{w}}  <-1/3$, would be  responsible for the present accelerated expansion of the Universe. Since the cosmological constant $ \Lambda$, added originally by Einstein in $1917$ to the field equations  of general relativity to make room for  a static universe,  acts as a homogeneous and isotropic source with $p/\rho=-1$, it reappeared  in this scenario as the most attractive and  natural candidate to describe  dark energy, receiving a great deal of attention from  cosmologists \cite{Peebles, Padmanabhan}. In fact, the currently available observational data  are consistent with the value $p/\rho=\textit{\texttt{w}
}=-1$ \cite{RiessII}, justifying the enormous theoretical interest in cosmological models with a $\Lambda$-term. However, the cosmological constant suffers from a serious problem: its theoretical value, estimated from particle physics, differs from the value compatible with  cosmological observations by  $30$ orders of magnitude in energy scale \cite{Trodden}. Thus, even if the sped up expansion can be phenomenologically well described by a cosmological constant, a  theoretical problem remains: how to explain the enormous discrepancy  between theory and experiment. Therefore, in spite of the possible harmony between observations and a cosmological term, we should have in mind that the cosmological term is little more than a mathematical device until one finds a more fundamental  understanding of the mechanism  by means of which the Universe accelerates.

In this work, a quantum flat FRW cosmological model  with nonrelativistic matter (dust) and a cosmological constant is studied. Since the classical model presents singularities the quantum approach seems adequate, for one of the main motivations of quantum cosmology is to investigate whether quantum-gravitational effects can  prevent singularities that appear in the classical theory. Moreover, there are some examples of quantum cosmological models that present quantum behavior for large scale factor  \cite{Kowalski, Colistete, Lemos}, which makes us suspect that in certain models  quantum effects may  be important during the entire history of the Universe and not only during a certain phase of its evolution as, for example, close to the singularities. Accordingly,  we are particularly interested in investigating if quantum effects can lead to significant modifications on  the late-time dynamics of the Universe.

This work is organized as follows. In section II we summarize the main aspects of the canonical formalism  which will be used in the accomplishment of our work. In section III the FRW minisuperspace is constructed. In section IV the    classical model is solved. In section V the model is quantized, and solutions to the Wheeler-DeWitt equation are found;
the expectation value of the scale factor at any time is calculated and its behavior spelled out. In section VI we present our final comments. Throughout,  units have been chosen such that $c=\hbar=16\,\pi\,G=1$.

\section{The Canonical Formalism}

In the canonical formalism, the line element is written as \cite{Arnowitt}
\begin{equation}
\label{eq.1}
ds^{2}=(-N^{2}+N_{i}N^{i})dt^{2}+2N_{i}dx^{i}dt+h_{ij}dx^{i}dx^{j},
\end{equation}
where $N$ is the lapse function, $N^{i}$ the displacement vector and $h_ {ij}$ the 3-metric induced on the spatial section $\Sigma_{t}$. Denoting by $\varphi_a$ the matter fields, the action for gravity (including the cosmological term) and its sources is given by
\begin{equation}
\label{eq.2}
\mathcal{S}=\int dt\int d^{3}x\,N\sqrt{h}[\,L_{g}-2\Lambda+L_{m}(\varphi_{a},\partial_{\mu}\varphi)]
\end{equation}
where $h$ is the determinant of the metric induced  on $\Sigma_t$, $L_{m}(\varphi_{}, \partial_{\mu}\varphi)$ is the matter Lagrangian and
\begin{equation}
\label{eq.3}
L_{g}=K_{ij}K^{ij}-K^{2}+\,^{3}\!R
\end{equation}
is the usual geometrodynamics  Lagrangian. Here, $K=K_{i}^{\ i}$ is the trace of the extrinsic curvature tensor $K_ {ij}$ and $^{3}\!R$ the Ricci scalar calculated with the metric of $\Sigma_{t}$. $K_ {ij}$ measures the curvature of the hypersurface $\Sigma_{t}$ with respect  to the higher dimensional space in which it is immersed and is given by
\begin{equation}
\label{eq.4}
K_{ij}=\frac{1}{2N}(N_{i|j}+N_{j|i}-\dot{h}_{ij}),
\end{equation}
where the vertical bar denotes covariant derivative calculated with the induced metric on $\Sigma_t$, and the dot indicates a partial derivative with respect to time.

Denoting by $\pi^{ij}$ and $\pi^{a}$ the momenta conjugate to $h_{ij}$ and $\varphi_{a}$, respectively, the action (\ref{eq.2}) can be written as
\begin{eqnarray}
\label{eq.5}
\mathcal{S}=\int dt\int d^{3}x\,(\pi^{ij}\dot{h}_{ij}+\pi^{a}\dot{\varphi}_{a}
-N\mathcal{H}^0-N_{i}\mathcal{H}^{i}),
\end{eqnarray} 
where
\begin{equation}
\label{eq.6}
\mathcal{H}^{0}=G_{ijkl}\pi^{ij}\pi^{kl}-\sqrt{h}\,^{3}\!R+2\sqrt{h}\,\Lambda
+\mathcal{H}^0_m
\end{equation}
and
\begin{equation}
\label{eq.7}
\mathcal{H}^{i}=2\pi^{ij}_{\,\,\,\,\vert j}+\mathcal{H}^i_m.
\end{equation}
Here, the Hamiltonian densities $\mathcal{H}^0_m$ and $\mathcal{H}^i_m$ are the matter contributions to  $\mathcal{H}^0$ and $\mathcal{H} ^i$, respectively, and
\begin{equation}
\label{eq.8}
G_{ijkl}=\frac{1}{2\sqrt{h}}(h_{ik}h_{jl}+h_{il}h_{jk}-h_{ij}h_{kl})
\end{equation}
is the supermetric or DeWitt metric \cite{DeWitt}. In (\ref{eq.5}), the lapse function and the displacement vector act as Lagrange multipliers leading to the constraints
\begin{equation}
\label{eq.9}
\mathcal{H}^0=0\,\,\mbox{and}\,\, \mathcal{H}^{i}=0.
\end{equation}
Finally, effecting the usual quantum prescriptions $\pi^{ij}\to\,- i\delta/\delta\,h_{ij}$, $\pi^{a}\to\,-i\delta/\delta\varphi_{a}$ and replacing the first constraint of (\ref{eq.9}) by the condition $\hat{\cal{H}}^{0}\Psi[h_{ij},\varphi_{a}]=0$, where $\hat{\cal{H}}^0$ is the operator associated with the super-Hamiltonian ${\cal{H}}^0$, we are led to the Wheeler-DeWitt  equation
\begin{equation}
\label{eq.10}
(-G_{ijkl}\frac{\delta}{\delta\,h_{ij}}\frac{\delta}{\delta\,h_{kl}}-\sqrt{h}\,^{3}\!R +2\sqrt{h}\,\Lambda+{\hat{\mathcal{H}}^0_m})\Psi=0\, ,
\end{equation}
which gives the dynamics of the quantum theory. In the above equation $\hat{\cal{H}}^0_m$ is the operator associated  with ${\cal{H}}^0_m$.

For the description of  the material content of the Universe we will use the Schutz formalism \cite{Schutz} for relativistic perfect fluids in which the 4-velocity of the fluid particles is written in terms of five potentials in the form
\begin{equation}
\label{eq.11}
u_{\nu}=\frac{1}{\mu}(\phi_{,\nu}+\alpha\beta_{,\nu}+\theta s_{,\nu}),
\end{equation}
where $s$ is the specific entropy and $\mu$ is the specific enthalpy, which is determined in terms of the other five potentials by the normalization condition $u^{\nu}u_{\nu}=-1$. The remaining potentials $\alpha,\ \beta,\ \theta$ and $\phi$ have no clear physical meaning. One of the advantages of the velocity-potential formalism is that dynamical degrees of freedom are  attributed to the fluid, from which a time variable can be naturally identified. In the Schutz formalism, the matter Lagrangian can be written as
\begin{equation}
\label{eq.12}
L_{m}=p,
\end{equation}
where $p$ is the fluid pressure. The matter Hamiltonian density is given by 
\begin{equation}
\label{eq.13}
{\mathcal{H}}^0_m=-\sqrt{h}\,T^{0}_{\,\,0},
\end{equation}
where $T^{0}_{\,\,0}$ is the time-time component of the  fluid energy-momentum tensor
\begin{equation}
\label{eq.14}
T^{\mu}_{\,\,\nu}=(\rho+p)u^{\mu}u_{\nu}+p\delta^{\mu}_{\nu}.
\end{equation}

\section{Minisuperspace Model}

The Wheeler-DeWitt equation is a second-order functional differential equation defined on the  infinite-dimensional space of all  possible configurations of spatial geometry and  matter fields, called superspace. This means that the Wheeler-DeWitt equation must be solved at each point $(x)$ of $\Sigma_t$ , which we are unable to do with the presently available mathematical techniques. However, interesting solutions can be obtained in the quantum cosmology context, where the infinitely many degrees of freedom of superspace are reduced to a finite number by means of symmetry arguments. The subspace thus constructed is called minisuperspace. Homogeneous universes are simple examples of minisuperspace models. Here we restrict our attention to a homogeneous, isotropic and spatially flat universe, whose metric is defined by the flat FRW line element
\begin{equation}
\label{eq.15}
ds^{2}=-N^{2}(t)dt^{2}+a^{2}(t)\delta_{ij}dx^{i}dx^{j},
\end{equation}
where $a(t)$ is the scale factor.
% of the universe and $\delta_{ij}=diag(1,1,1)$ is the Kronecker delta.

To complete our model
%, still lacks specify completely the fluid movement. That is made supplying its EoS, $p=p(\rho)$.
we assume that the cosmic fluid satisfies a barotropic equation of state $p/\rho=\textit{\texttt{w}}$ with $\textit{\texttt{w}}=\mbox{constant}$. The symmetries of the line element (\ref{eq.15}) reduce the super-Hamiltonian (\ref {eq.6}) to the form \cite{Alvarenga}
\begin{equation}
\label{eq.16}
\mathcal{H}^{0}=-\frac{p_{a}^{2}}{24a}+2\Lambda\,a^{3}+a^{-3\textit{\texttt{w}}}p_{\phi}^{1+\textit{\texttt{w}}}e^{s},
\end{equation}
where $p_{a}$ and $p_{\phi}$ are the  momenta conjugate to the scale factor $a$ and the potential $\phi$, respectively. Finally, performing the canonical transformation $(a,\phi,s,p_a,p_{\phi},p_s)\to(a,\psi ,\eta,p_a,p_{\psi},p_{\eta})$ defined by

\begin{equation}
\label{eq.17a}
\begin{array}{ll}
\eta=-p_s\,p_{\phi}^{-(1+\textit{\texttt{w}}\,)}\,e^{-s}\, ,\,\, p_{\eta}=p_{\phi}^{1+\textit{\texttt{w}}}\,e^s\, \\
\displaystyle \psi=\phi + (\textit{\texttt{w}}+1)\frac{p_s}{p_{\phi}}\, ,\,\, p_{\psi}=p_{\phi}
\end{array}
\end{equation}

%\begin{eqnarray}
%\eta=-p_s\,p_{\phi}^{-(1+w)}\,e^{-s}\, ,\,\, p_{\eta}=p_{\phi}^{1+w}\,e^s\, ,\nonumber\\
%\psi=\phi + (w+1)\frac{p_s}{p_{\phi}}\, ,\,\, p_{\psi}=p_{\phi}
%\label{eq.17}
%\end{eqnarray}
we get
\begin{equation}
\label{eq.18}
{\mathcal{H}}^0=-\frac{p_{a}^{2}}{24a}+2\Lambda a^{3}+a^{-3\textit{\texttt{w}}}p_{\eta}.
\end{equation}

\section{Classical Model}

From now on we shall restrict ourselves to the case $\textit{\texttt{w}}=0$ (dust). Then the super-Hamiltonian (\ref{eq.18}) reduces to
\begin{equation}
\label{eq.19}
{\mathcal{H}}^0=-\frac{p_{a}^{2}}{24a}+2\Lambda a^{3}+p_{\eta}.
\end{equation}
In the cosmic-time gauge, $N=1$, the classical equations of motion are given by
\begin{eqnarray}
\label{eq.20}
\dot{a}=\frac{\partial\mathcal{H}^{0}}{\partial p_a}=-\frac{p_a}{12 a},\,\,\, \dot{p}_a=-\frac{\partial\mathcal{H}^{0}}{\partial a}=-\frac{p_{a}^{2}}{24 a^2}-6\Lambda a^2\\
\label{eq.21}
\dot{\eta}=\frac{\partial\mathcal{H}^{0}}{\partial p_{\eta}}=1,\,\,\,\dot{p}_{\eta}=-\frac{\partial\mathcal{H}^{0}}{\partial \eta}=0.\,\,\,\,\,\,\,\,\,\,\,\,\,
\end{eqnarray}
The first of equations (\ref{eq.21}) supplies the relation between $\eta$ and the cosmic time $t$, the second equation of (\ref{eq.21}) says that $p_{\eta}$ is a constant and the first of equations (\ref{eq.20}) combined with the constraint ${\cal{H}}^0=0$ lead to the Friedmann equation
\begin{equation}
\label{eq.22}
\big(\frac{\dot{a}}{a}\big)^{2}-\frac{1}{3}\Lambda=\frac{p_{\eta}}{6 a^{3}}.
\end{equation}
Comparing (\ref{eq.22}) with the usual form of the Friedmann equation, we find that $p_{\eta}=\rho_{0} a_{0}^{3}=\rho a^{3}$ is the energy  stored in the fluid ($\rho$ is the energy density of the  fluid and the subscript  zero corresponds to the present time  value of the quantity). Solving the Friedmann equation we get
\begin{equation}
\label{eq.23}
a^{3}(t)=\frac{p_{\eta}}{4\vert\Lambda\vert}\Bigl[1-\cos(\sqrt{3\vert\Lambda\vert}\,t)\Bigr]\,\,\,\mbox{if}\,\,\,\Lambda<0
\end{equation}
and
\begin{equation}
\label{eq.24}
a^{3}(t)=\frac{p_{\eta}}{4\Lambda}\Bigl[\cosh(\sqrt{3\Lambda}\,t)-1\Bigr]\,\,\,\mbox{if}\,\,\,\Lambda>0 \, .
\end{equation}
These solutions describe  an oscillating universe and an eternally expanding universe, respectively. Both  scenarios present  singularities. We now proceed to examine whether these singularities persist and what the evolution of the Universe looks like in the quantum regime.

\section{Quantization}

The quantization procedure is greatly simplified by first performing the canonical transformation  $(a,\eta,p_a,p_{\eta})\to(\xi,\eta,p,p_{\eta})$ defined by
\begin{equation}
\label{eq.25}
\xi=\frac{4}{\sqrt{3}}a^{3/2},\ \ p=\frac{1}{\sqrt{12}}a^{-1/2}p_{a},
\end{equation}
which allows us to write the super-Hamiltonian (\ref{eq.19}) in the much simpler form
\begin{equation}
\label{eq.26}
{\mathcal{H}}^0=-\frac{1}{2}p^{2}+\frac{3}{8}\Lambda\xi^{2}+p_{\eta}.
\end{equation}
Applying the usual quantum prescriptions $p\to\hat{p}=-i\partial/\partial\xi,\ p_{\eta}\to
\hat{p}_{\eta}=i\partial/\partial\tau$ ($\tau=-\eta$) and the requirement $\hat{\cal{H}}^0\Psi=0$, where $\hat{\cal{H}}^0$ is the  operator corresponding to the super-Hamiltonian (\ref{eq.26}), the Wheeler-DeWitt equation in mini-superspace becomes
\begin{equation}
\label{eq.27}
\hat{H}\Psi=i\frac{\partial\Psi}{\partial\tau},
\end{equation}
with the Hamiltonian operator $\hat{H}$  given by
\begin{equation}
\label{eq.28}
\hat{H}=-\frac{1}{2}\frac{\partial^{2}}{\partial\xi^{2}}-\frac{3}{8}\Lambda\xi^{2}.
\end{equation}
Since $\xi \geq 0$,   solving the Wheeler-DeWitt equation (\ref{eq.27}) in minisuperspace is tantamount to solving the Schr\"odinger equation for a particle of mass $m=1$ subjected to the potential
\begin{equation}
\label{eq.29}
V(\xi)=
\left\{
\begin{array}{ll}
\infty\ &  \mbox{if} \ \xi<0 \\
-3\Lambda\xi^2/8 &  \mbox{if} \ \xi\geq0.\\
\end{array}
\right.
\end{equation}
The case $\Lambda<0$ corresponds to a harmonic oscillator of frequency $\omega=\sqrt{3\,\vert\Lambda\vert}/2$, whereas the case $\Lambda>0$ corresponds to
%the quantum mechanics of a particle in
an inverted oscillator potential, which can formally be obtained from the usual oscillator potential by the replacement  $\omega\to i\omega$ \cite{Barton}. The infinite barrier at $\xi=0$
reflects the condition that the scale factor $a$ cannot be negative.

Incidentally, although we have chosen the Dirac quantization method, this does not impair the generality of our treatment as far as quantization is concerned. Because the super-Hamiltonian
(\ref{eq.26})  is linear in the momentum $p_{\eta}$, it is well known that Dirac quantization and canonical quantization after a full phase-space reduction  yield the same physics.

%The referee asks us to comment ``whether fully reduced quantization is expected to yield the same physics" as  the ``Dirac quantization" that we chose. As he himself remarks, the answer  is easy in our case: since the super-Hamiltonian
%(26) of the revised manuscript is linear in the momentum $p_{\eta}$, it is well known that Dirac quantization and  quantization preceded by  phase-space reduction  yield the same physics. We have added a comment to this effect right after the paragraph of the  revised manuscript that contains Eq. (29).

On $L^2(0,\infty )$ the self-adjointness of the Hamiltonian operator requires that the wave functions be restricted by the condition \cite{Reed}
\begin{equation}
\label{eq.30}
\Psi^{\prime}(0,\tau)=\gamma\Psi(0,\tau)
\end{equation}
where $\Psi^{\prime}=\partial\Psi/\partial\xi$ and $\gamma\,\epsilon\,(-\infty,\,\infty]$.
%Owing to form of the potential (\ref{eq.29}),
%For the sake of simplicity, we will consider only the case in which $\Psi(0,\tau)=0$, that is, $\gamma=\infty$.
Although stationary solutions can be constructed, these are not of  interest since we know that we live in an expanding Universe. Therefore, we must find wave functions  that describe an evolving Universe. To this end,  we need the propagator for the Schr\"odinger equation (\ref{eq.27}) in the restricted Hilbert space $L^2(0,\infty)$. 
%Since the obtaining of the propagator for an arbitrary $\gamma$ is a very hard task, 
To our knowledge, this propagator is not known for  arbitrary $\gamma$. Therefore, 
here we limit our discussion to the cases $\Psi(0,\tau)=0$ and $\Psi^{\prime}(0,\tau)=0$, that is, $\gamma=\infty$ and $\gamma=0$, respectively. In such cases the propagator is given, respectively, by \cite{Clark, Farhi}%For $\gamma=\infty$ the propagator  is given by \cite{Clark, Farhi}
\begin{equation}
\label{eq.31a}
G_{\infty}(\xi,\,\xi^{\prime},\,\tau)=G(\xi,\,\xi^{\prime},\,\tau)-G(\xi,\,-\xi^{\prime},\,\tau)
\end{equation}
and
\begin{equation}
\label{eq.31b}
G_{0}(\xi,\,\xi^{\prime},\,\tau)=G(\xi,\,\xi^{\prime},\,\tau)+G(\xi,\,-\xi^{\prime},\,\tau),
\end{equation}
where $G(\xi,\,\xi^{\prime},\,\tau)$ is the usual propagator in the standard Hilbert space $L^{2}(-\infty,\,\infty)$.

\subsection{The case $\Psi(0,\tau )=0$ ($\gamma=\infty$)}

For $\Lambda<0$, the Hamiltonian (\ref{eq.28}) refers to a harmonic oscillator of frequency $\omega=\sqrt{3\,\vert\Lambda\vert}/2$ and $m=1$, so that the usual propagator is given by
% and $G(\xi,\,\xi^{\prime},\,\tau)$ is given by
\begin{eqnarray}
\label{eq.32}
G(\xi,\,\xi^{\prime},\,\tau)&=&\Big[\frac{\omega}{2\pi i\, \sin\omega\tau}\Bigr]^{1/2}\exp\Bigl\{\frac{i\,\omega}{2\,\sin\omega\tau}\nonumber\\
&\times& \bigl[(\xi^{\prime2}+\xi^{2})\cos\omega\tau-2\xi^{\prime}\xi\bigr]\Bigr\}.
\end{eqnarray}

Let us take for  initial wave function
\begin{equation}
\label{eq.33}
\Psi_{o}(\xi,\,0)=\Big[\frac{16(\alpha\omega)^{3}}{\pi}\Big]^{1/4}\,\xi\,e^{-\frac{\omega}{2}(\alpha+i\,\beta)\xi^2},\ \alpha>0\, ,
\end{equation}
where the real parameters $\alpha$ and $\beta$ are left arbitrary in order to allow a certain freedom of choice within the restricted class of Gaussian initial wave functions.
The wave function at any time  is
\begin{eqnarray}
\label{eq.34}
\Psi_{o}(\xi,\,\tau)&=&\int_{0}^{\infty}d\xi^{\prime}G_{\infty}(\xi,\,\xi^{\prime},\,\tau)\Psi_{o}(\xi^{\prime},0)\nonumber\\
&=&\int_{-\infty}^{\infty}d\xi^{\prime}G(\xi,\,\xi^{\prime},\,\tau)\Psi_{o}(\xi^{\prime},0),
\end{eqnarray}
where we have taken advantage of the fact that $\Psi_o(\xi,\,0)$ is an odd function of $\xi$ to extend the integration to the whole real line. Inserting  (\ref{eq.32}) and (\ref{eq.33}) in (\ref{eq.34}) we obtain
\begin{equation}
\label{eq.35}
\Psi_{o}(\xi,\,\tau)=\Big[\frac{16(\alpha\,\omega)^{3}}{\pi\,z^{6}(\tau)}\Big]^{1/4}\xi\,\exp\Big[\frac{i\,\omega}{2}\frac{z(\tau+\frac{\pi}{2\omega})}{z(\tau)}\xi^{2}\Big],
\end{equation}
where
\begin{equation}
\label{eq.36}
z(\tau)=\cos\omega\tau-\beta\sin\omega\tau+i\alpha\,\sin\omega\tau.
\end{equation}

Now, by (\ref{eq.25}) we have that $a=(3/16)^{1/3}\xi^{2/3}$, so that the expectation value of the scale factor is
\begin{eqnarray}
\label{eq.37}
\langle a\rangle_{o}(\tau)&=&\Big(\frac{3}{16}\Big)^{1/3}\int_{0}^{\infty}d\xi\Psi_{o}(\xi,\tau)\xi^{2/3}\Psi_{o}^{*}(\xi,\tau)\nonumber\\
&=&\frac{2}{\sqrt{\pi}}\Big(\frac{3}{16\,\omega\,\alpha}\Big)^{1/3}\Gamma\Big(\frac{11}{6}\Big)\vert z(\tau)\vert^{2/3}
\end{eqnarray}
or
\begin{eqnarray}
\label{eq.38}
\langle a\rangle_{o}^{3}(\tau)&=&\frac{3}{2\,\alpha\,\omega\,\pi^{3/2}}\,\Gamma^3\Big(\frac{11}{6}\Big)
\big[\cos^2\omega\tau\nonumber\\
&+&(\alpha^2+\beta^2)\sin^2\omega\tau-\beta\,\sin2\omega\tau\big].
\end{eqnarray}
It can be  clearly seen from (\ref{eq.36}) and (\ref{eq.37}) that $\langle a\rangle_o$ never vanishes, indicating the absence of singularities in the quantum theory.  By (\ref{eq.21}), we have that the time $\tau$ and the cosmic time $t$ are related by $\tau=T-t$, where $T$ is an integration constant. Remembering that $\omega=\sqrt {3\vert\Lambda\vert}/2$ and choosing $2\omega\,T=\tan^{-1}[2\beta/(\alpha^2+\beta^2-1)]$ so that $\langle a\rangle_{o}$ is minimum when $t=0$,  equation (\ref{eq.38}) takes the form

\begin{eqnarray}
\label{eq.42}
\langle a\rangle_{o}^{3}(t)&=&\frac{3(1+\alpha^2+\beta^2)}{2\,\alpha\,\sqrt{3\,\pi^{3}\,\vert\Lambda\vert}}\,\Gamma^3\Big(\frac{11}{6}\Big)\nonumber\\
&\times&\Big[1-\frac{\sqrt{(1-\alpha^2-\beta^2)^2+4\beta^2}}{1+\alpha^2+\beta^2}\,\cos(\sqrt{3\vert\Lambda\vert}\,t)\Big].\nonumber\\
\end{eqnarray}
Comparing this equation with (\ref{eq.23}) we see that % the quantization affects the  universe's period  of  oscillation: the quantum period is longer than the classical one by the factor $\sqrt{3/2}$. Moreover,
$\alpha$ can be considered a measure of how nearly ``singular''  the universe is at $t=0$, in the sense that the  closer  $\alpha$ is to zero the  denser and hotter the universe  was at $t=0$. 
%Since $\lim_{\alpha \to 0} \vert \Psi_o(\xi ,0 )\vert^2 = \delta (\xi )$, the initial singularity $\langle a\rangle_o=0$ is enforced at $t=0$ and the classical limit is attained for $\alpha \to 0$. If $\beta =0$ and $\alpha =1$, $\langle a\rangle_o$ is constant. In this case the wave function (\ref{eq.35}) becomes time independent and  may be considered as describing a static universe.
For $\beta=0$ and $\alpha=1$, $\langle a \rangle_{o}$ is constant and the squared modulus of the wave function (\ref{eq.35}) is time independent, so that in this case (\ref{eq.35}) may be considered as describing a static universe.

%The fact that the period of oscillation of the universe in the quantum regime is different from the classical one indicates that quantum effects may influence the entire history of the universe.
%Since the cosmological constant has dimension of $(\mbox{time})^{-2}$,
%it is easy to check that the only combination of $c, G, \hbar$ and $\Lambda$ with the dimensions of inverse time squared is $c^2\Lambda$. Therefore,
%the change in the period of oscillation of the Universe induced by the quantization process does not depend on the Planck constant. This is a rare event, reminiscent of the Rutherford scattering cross section in quantum mechanics, whose value does not depend on $\hbar$ either.

For $\Lambda>0$, the  wave function $\Psi(\xi,\tau)$ can be obtained by making $\omega\to i\omega$ in (\ref{eq.32}) and using the identities $ \cos ix=\cosh x$ and $\sin ix=i\sinh x$. Proceeding in this fashion, the propagator for the anti-oscillator is found to be

\begin{eqnarray}
\label{eq.43}
G(\xi,\,\xi^{\prime},\,\tau)&=&\big[\frac{\omega}{2\pi i\, \sinh(\omega\tau)}\big]^{1/2}\exp\Big\{\frac{i\,\omega}{2\sinh(\omega\tau)}\nonumber\\
&\times& \big[(\xi^{\prime2}+\xi^{2})\cosh(\omega\tau)-2\xi^{\prime}\xi\big]\Big\}.
\end{eqnarray}
Picking the same initial wave function (\ref{eq.33}), equation (\ref{eq.34}) now yields
\begin{equation}
\label{eq.44}
\Psi_o(\xi,\,\tau)=\Big[\frac{16(\alpha\,\omega)^{3}}{\pi\,\zeta^{6}(\tau)}\Big]^{1/4}\xi\,\exp\Big[\frac{\omega}{2}\frac{\zeta(\tau+i\,\frac{\pi}{2\omega})}{\zeta(\tau)}\xi^{2}\Big],
\end{equation}
where
\begin{equation}
\label{eq.45}
\zeta(\tau)=\cosh\omega\tau-\beta\,\sinh\omega\tau+i\,\alpha\,\sinh\omega\tau.
\end{equation}
In this case, the expectation value of the scale factor is
\begin{eqnarray}
\label{eq.46}
\langle a\rangle_{o}(\tau)=\frac{2}{\sqrt{\pi}}\Big(\frac{3}{16\,\omega\,\alpha}\Big)^{1/3}\Gamma\Big(\frac{11}{6}\Big)\vert \zeta(\tau)\vert^{2/3}
\end{eqnarray}
or
\begin{eqnarray}
\label{eq.47}
\langle a\rangle_o^3(\tau)&=&\frac{3}{2\,\alpha\,\omega\,\pi^{3/2}}\,\Gamma^3\Big(\frac{11}{6}\Big)
\big[\cosh^2(\omega\,\tau)\nonumber\\
&+&(\alpha^2+\beta^2)\sinh^2(\omega\,\tau)-\beta\,\sinh(2\,\omega\,\tau)\big],
\end{eqnarray}
which is never  zero. %As in the previous case, the condition for  the expectation value of the energy to be time independent is that $\alpha$ and $\beta$ satisfy the constraint (\ref{eq.40}), in which case the expectation value of the energy vanishes:
%\begin{equation}
%\label{eq.48}
%\langle \hat{H}\rangle=0.
%\end{equation}
In terms of the cosmic time, making $2\omega\,T=\tanh^{-1}[2\beta/(1+\alpha^2+\beta^2)]$, so that $d\langle a\rangle/dt=0$ at $t=0 $, the expectation value of the scale factor becomes
\begin{eqnarray}
\label{eq.49}
\langle a\rangle_{o}^{3}(t)&=&\frac{3(\alpha^2+\beta^2-1)}{2\,\alpha\,\sqrt{3\,\pi^{3}\,\vert\Lambda\vert}}\,\Gamma^3\Big(\frac{11}{6}\Big)\nonumber\\ 
&\times&\Big[\sqrt{\frac{(1+\alpha^2+\beta^2)^2-4\beta^2}{(\alpha^2+\beta^2-1)^2}}\,\cosh(\sqrt{3\Lambda}\,t)-1\Big].\nonumber\\
\end{eqnarray}
%A comparison of this result with the classical solution (\ref{eq.24}) shows  that the present expansion rate is reduced in the quantum case, that is, quantum effects influence the late-time dynamics of the universe. The previous remark applies: this effect does not depend on the Planck constant.
We assume  $\alpha^2+\beta^2 > 1$ in order that the second time derivative of $\langle a\rangle_{o}(t)$ may have zeroes, as will be discussed below. Note that the expectation value of the scale factor bounces at $t=0$, when its value is as small as possible. Again, the  closer  $\alpha$ is to zero the  denser and hotter--- or nearly singular --- the universe  was at the bounce.

For $\Lambda>0$, the influence of the quantum gravitational effects on the late-time dynamics of the universe can be revealed by  evaluating the times $t_{*}$ at which  the acceleration changes sign  and comparing the classical  values with those obtained from the quantum model.  The times $t_{*}$ are the zeroes of the second derivative of the scale factor, in the classical theory,  or of its expectation value, in the quantum theory. For a function $f$ of the form $f(t)=g(t)^{1/3}$ one has ${\ddot f}=0$ only if $3 g {\ddot g}=2{\dot g}^2$.
From (\ref{eq.24}) the classical theory furnishes a single time
\begin{equation}
\label{ClassicTansitionTime}
t_{*}^{c}=\frac{1}{\sqrt{3\,\Lambda}}\cosh^{-1}2\, ,
\end{equation}
whereas from (\ref{eq.49}) the  quantum theory yields the two times
\begin{equation}
\label{QuantumTansitionTime}
t_{*\pm}^{q}=\frac{1}{\sqrt{3\,\Lambda}}\cosh^{-1}\Big( \frac{3\pm\sqrt{9-8\,\kappa^2}}{2\,\kappa}\Big),
\end{equation}
where 
\begin{equation}
\label{kparameter}
\kappa=\sqrt{\frac{(1+\alpha^2+\beta^2)^2-4\beta^2}{(\alpha^2+\beta^2-1)^2}}.
\end{equation}
Since $t_{*\pm}^{q}$ are real numbers and $\alpha>0$ by definition, we have that $1<\kappa\leq 3/2\,\sqrt{2}$. Thus, the times $t_{*-}^{q}$ and $t_{*+}^{q}$ lie, respectively, in the ranges
\begin{equation}
\label{ConstraintOnTqMinus}
0<t_{*-}^{q}\leq\frac{1}{\sqrt{3\,\Lambda}}\cosh^{-1}\sqrt{2}
\end{equation}
and
\begin{equation}
\label{ConstraintOnTqPlus}
\frac{1}{\sqrt{3\,\Lambda}}\cosh^{-1}\sqrt{2}\leq t_{*+}^{q}<t_{*}^{c}.
\end{equation}
The time $t_{*-}^q$ marks the end of an early accelerated expansion phase started right after the bounce and the beginning of a phase dominated by nonrelativistic matter. The time $t_{*+}^q$ represents the beginning of the present phase of accelerated expansion of the Universe.
%If $\kappa=3/2\,\sqrt{2}$ then $t_{*-}^q=t_{*+}^q$ and there is no early accelerated expansion phase.
Our quantum model entails that the  Universe's present phase of accelerated expansion should have started earlier than predicted by the classical theory, and suggests that a single cosmological constant can account for both an early inflationary phase and the present accelerated expansion  of the Universe.

%Therefore, the quantum theory predicts that the  universe's present phase of accelerated expansion started earlier than predicted by the classical theory. 
In order to give an estimate  of the times $t_{*\pm}^q$ we write $\Lambda=3\,H_{0}^{2}\,\Omega_{\Lambda}$ and take $\Omega_{\Lambda}=0.7$, as indicated by the current observational data. Thus, we have that the time $t_{*-}^q$ lies in the range
\begin{equation}
\label{ConstraitOnTqObservational1}
0< t_{*-}^{q}\leq 0.35\,H_{0}^{-1}
\end{equation}
and the time $t_{*+}^q$ lies in the range
\begin{equation}
\label{ConstraitOnTqObservational}
0.35\,H_{0}^{-1}\leq t_{*+}^{q}<0.52\,H_{0}^{-1}\, ,
\end{equation}
that is, the early accelerated expansion phase  occurred between $14$ and $9$ billion years ago, and the present accelerated expansion began between about $9$ and $7$ billion years ago.

%Taking the negative sign in (\ref{QuantumTansitionTime}) we find that $t_{*-}^{q}$ lies in the range
%This instant $ t_{*-}^{q}$ may be interpreted as marking the end of an inflationary phase which started right after the bounce, and the beginning of a phase dominated by nonrelativistic matter. It is remarkable that a single cosmological constant can account for an early inflationary phase and the present accelerated expansion regime of the Universe, with a decelerated phase in between, during  which nonrelativistic matter dominated.

The above results hold true only if $\alpha^2 + \beta^2 > 1$, that is, for a subset of the  class of initial wave functions (\ref{eq.33}). If 
$\alpha^2 + \beta^2 \leq 1$ the expectation value of the scale factor describes a forever accelerated expansion from the bounce at $t=0$. This suggests that observational data on the relatively recent evolution of the Universe may shed some light on what the initial quantum state of the Universe might have been.

\subsection{The case $\Psi^{\prime}(0,\tau)=0$ $(\gamma=0)$}

Now, in order to have a taste of  the problem whether the choice of a different self-adjoint extension of the Hamiltonian operator  can modify our results, we take the initial state described  by the normalized wave function
\begin{equation}
\label{BoundaryCondition2}
\Psi_{e}(\xi,\,0)=\Big(\frac{4\alpha\omega}{\pi}\Big)^{1/4}\,e^{-\frac{\omega}{2}(\alpha+i\,\beta)\xi^2},\ \alpha>0\, ,
\end{equation}
which satisfies the boundary condition (\ref{eq.30}) with $\gamma=0$. Since $\Psi_{e}(\xi,\,0)$ is  an even function of $\xi$, the wave function at any time is given by
\begin{eqnarray}
\label{TimeDependentWaveFunction1}
\Psi_{e}(\xi,\,\tau)&=&\int_{0}^{\infty}d\xi^{\prime}G_{0}(\xi,\,\xi^{\prime},\,\tau)\Psi_{e}(\xi^{\prime},0)\nonumber\\
&=&\int_{-\infty}^{\infty}d\xi^{\prime}G(\xi,\,\xi^{\prime},\,\tau)\Psi_{e}(\xi^{\prime},0).
\end{eqnarray}

For a negative cosmological constant, we have
\begin{equation}
\label{TimeDependentWaveFunction2}
\Psi_{e}(\xi,\,\tau)=\Big[\frac{4\alpha\,\omega}{\pi\,z^{2}(\tau)}\Big]^{1/4}\,\exp\Big[\frac{i\,\omega}{2}\frac{z(\tau+\frac{\pi}{2\omega})}{z(\tau)}\xi^{2}\Big],
\end{equation}
where $z(\tau)$ is given as before by (\ref{eq.36}). In this case the expectation value of the scale factor is 
\begin{equation}
\label{ScaleFactorEvenCC<0}
\langle a\rangle_{e}(\tau )=\frac{3}{5}\,\langle a\rangle_{o}(\tau )
\end{equation}
with $\langle a\rangle_{o}(\tau )$ given by (\ref{eq.38}).

For $\Lambda>0$, the wave function at any time is
\begin{equation}
\label{TimeDependentWaveFunction3}
\Psi_{e}(\xi,\,\tau)=\Big[\frac{4\alpha\,\omega}{\pi\,\zeta^{2}(\tau)}\Big]^{1/4}\,\exp\Big[\frac{i\,\omega}{2}\frac{\zeta(\tau+\frac{\pi}{2\omega})}{\zeta(\tau)}\xi^{2}\Big],
\end{equation}
where $\zeta(\tau)$ is given by (\ref{eq.45}). The expectation value of the scale factor is
\begin{equation}
\label{ScaleFactorEvenCC>0}
\langle a\rangle_{e}(\tau )=\frac{3}{5}\,\langle a\rangle_{o}(\tau )
\end{equation}
with $\langle a\rangle_{o}(\tau )$ given by (\ref{eq.47}). 

This last result means that our previous conclusions concerning the case $\Lambda>0$ remain the same. 

\section{Conclusions and Outlook}

We have shown how to obtain exact solutions to the   Wheeler-DeWitt equation for a flat FRW universe with dust and a cosmological constant. In the case of negative cosmological constant, the classical scale factor evolves from a singular beginning (Big Bang) to a singular end (Big Crunch). Quantization leads to  a nonsingular and forever oscillating model, since the expectation value of the scale factor never vanishes. For positive cosmological constant, the classical scale factor grows eternally from a singular beginning. Quantization leads to  a nonsingular and forever expanding universe after a bounce. 

%In both cases the quantization process affects the expansion rate of the universe as compared to the corresponding %classical result, but the change does not depend on the Planck constant. This is unexpected, since it occurs only very %rarely, as in the case  of the Rutherford scattering cross section in quantum mechanics, whose value does not depend %on $\hbar$. In the case of positive   cosmolog
%ical constant,
%quantization modifies the present expansion rate of the Universe, suggesting that  quantum effects may not be %important only close to singularities, but may influence the late-time  dynamics of the Universe.

In the more realistic case of positive $\Lambda$, it is remarkable that a single cosmological constant can give rise to an early inflationary phase and a late regime of accelerated expansion, with a period of decelerated expansion in between. Considering the crudeness of our model, this result gives reason for hope  that a more sophisticated model, including radiation, may be able to provide a more accurate description of the several phases the Universe is believed to have gone through.

There are important issues that deserve further scrutiny. Do other self-adjoint extensions of the 
Hamiltonian operator, that is, other values of $\gamma$ in Eq.(\ref{eq.30}), qualitatively change our results? We have checked that the choices $\gamma =0$ and $\gamma =\infty $ lead to the same results, but this is a long way from proving that the results persist no matter what the value chosen for $\gamma$. The lack of a closed expression for the harmonic oscillator propagator on the half-line for arbitrary $\gamma$ is an obstacle to be overcome if one wishes to give an unequivocal answer to this question. It should not go unnoticed that
the estimates (\ref{ConstraitOnTqObservational1}) and (\ref{ConstraitOnTqObservational})  rely on the classical  value of the cosmological constant. Is a  fully quantum treatment, including a reasonable quantum estimate for $\Lambda$, feasible?
%Perhaps this can be done --- since for a positive  cosmological constant  the Hamiltonian operator is that of an anti-oscillator --- by evaluating the time of persistence of the universe
%For a positive  cosmological constant  we calculate the probability of the universe to be found
%near the unstable equilibrium position $a=0$
%of an anti-oscillator 
%\cite{Barton}.
%-region  and the time of persistence of the universe in this region ...
It is also worth investigating whether a more fundamental description of the matter content, by means of fields, affects the results reported here. We expect to take up at least some of these issues in a future work.

\vspace{.5cm}

\begin{acknowledgments}
Ed\'esio M. Barboza Jr. is supported by Conselho Nacional de Desenvolvimento Cient\'ifico e Tecnol\'ogico (CNPq), Brazil.
\end{acknowledgments}


\begin{thebibliography}{99}
\bibitem{Riess} Riess,  A. G. {\it et al.}, Astron. J. {\bf 116}, 1009 (1998).
\bibitem{Perlmutter} Perlmutter, S. {\it et al.}, Astrophys. J. {\bf 517}, 565 (1999).
\bibitem{Peebles} Peebles, P. J. E. and  Ratra, B.,  Rev. Mod. Phys. {\bf 75}, 559 (2003).
\bibitem{Padmanabhan} Padmanabhan, T., Phys. Rept. {\bf 380}, 235 (2003).
\bibitem{RiessII} Riess, A. G.  {\it {\it et al.}}, Astrophys. J. {\bf 607}, 665 (2004).
%\bibitem{Weinberg}
\bibitem{Trodden} Trodden, M. and Carroll, S. M. {\it TASI Lectures: Introduction to Cosmology}, astro-ph/0401547.
\bibitem{Kowalski} Kowalski-Glikman, J. and Vink, J. C., Class. Quantum Grav. {\bf 7}, 901 (1990).
\bibitem{Colistete} Colistete Jr., R., Fabris, J. C. and Pinto-Neto, N., Phys. Rev. D {\bf 57}, 4707 (1998).
\bibitem{Lemos} Lemos N. A., and Alvarenga, F. G., Gen. Rel. Grav. {\bf 31}, 1743 (1999). 
\bibitem{Arnowitt} Arnowitt, R., Deser, S. and Misner, C. W. in
\textit{Gravitation: An Introduction to Current Research}, edited by Witten, L.  (Wiley, New York, 1962), 227-265 [gr-qc/0405109] .
%\bibitem{Wald} Wald, R. M., \textit{General Relativity} (The University of Chicago Press, Chicago, 1984).
\bibitem{DeWitt} DeWitt, B. S., Phys. Rev. \textbf{160}, 1113 (1967).
\bibitem{Schutz} Schutz, B. F.,  Phys. Rev. D \textbf{2}, 2762 (1970); Schutz, B. F.,  Phys. Rev. D \textbf{4}, 3559 (1971).
\bibitem{Alvarenga} Alvarenga, F. G., Fabris, J. C., Lemos, N. A. and Monerat, G. A., Gen. Rel. Grav.  \textbf{34}, 651 (2002).
\bibitem{Barton} Barton, G., Ann. Phys. \textbf{166}, 322 (1986).
\bibitem{Reed} Reed, I. M.
and Simon, B., {\it Methods of Modern Mathematical Physics Vol. II: Fourier Analysis, Self-Adjointness} (Academic, NY, 1975), Section X.1, Example 2.
\bibitem{Clark} Clark, T. E.,  Menikoff, R. and Sharp, D. H.,  Phys. Rev. D {\bf 22}, 3012 (1980).
\bibitem{Farhi}Farhi, E. and Gutmann, S., Int. J. Mod. Phys. A {\bf 5}, 3029 (1990).
\end{thebibliography}
\end{document}